\documentclass[a4paper, superscriptaddress, aps, prl, onecolumn, showkeys]{revtex4}
\usepackage{units}
\usepackage{graphicx}
\usepackage{amsmath}
\usepackage{amssymb}

\begin{document}

\author{Daniel Platz}
\email{platz@kth.se}
\author{Daniel Forchheimer}
\affiliation{Royal Institute of Technology (KTH), Section for Nanostructure Physics, Albanova University Center, SE-106 91 Stockholm, Sweden}
\author{Erik A. Thol\'{e}n}
\affiliation{Intermodulation Products AB, SE-169 58 Solna, Sweden}
\author{David B. Haviland}
\affiliation{Royal Institute of Technology (KTH), Section for Nanostructure Physics, Albanova University Center, SE-106 91 Stockholm, Sweden}

\title{Polynomial force approximations and multifrequency atomic force microscopy}

\begin{abstract}
We present polynomial force reconstruction from experimental intermodulation
atomic force microscopy (ImAFM) data. We study the tip-surface force
during a slow surface approach and compare the results with amplitude-dependence
force spectroscopy. Based on polynomial force reconstruction we generate
high-resolution surface property maps of polymer blend samples. The
polynomial method is described as a special example of a more general
approximataive force reconstruction, where the aim is to determine
model parameters which best approximate the measured force spectrum.
This approximative approach is not limited to spectral data and we
demonstrate how is can adapted to a force quadrature picture.
\end{abstract}

\maketitle

\section{Introduction}

The combination of high-resolution imaging\cite{Ohnesorge1993,Giessibl1995,Fukuma2005,Gan2009}
and high-accuracy force measurements\cite{Durig2000,Sader2005,Holscher2006,Lee2006,Hu2008,Katan2009}
is a strong driving force for the development of atomic force microscopy
(AFM). The advent of multifrequency AFM resulted in a variety of new
measurement techniques enabling enhanced contrast and spatial mapping
of surface properties on a wide range of samples\cite{Garcia2012}.
However, multifrequency AFM creates more data than conventional AFM,
which both complicates the interpretation of measurement results and
offers the possibility of much more detailed surface analysis. One
of the goals when interpreting AFM data is the reconstruction of the
force between a surface and the sharp tip at the end of the oscillating
cantilever, while scanning. This reconstruction is readily possible
by means of the Fourier transform if the tip's motion in response
to this force and the cantilever's linear response function are known
over a wide frequency band\cite{Durig2000a,Stark2002,Legleiter2006,Sahin2007}.
However, the tip motion is often only measurable in a narrow frequency
band around a cantilever resonance, since the cantilever transfer
function sharply attenuates other frequency components of the tip
motion, placing them below the detection noise floor. With this measured
partial motion spectrum, the original force cannot be recovered with
a simple Fourier transform and additional assumptions about the functional
representation of the tip-surface force are required.

These assumptions, which can be expressed with a finite set of parameters,
result in a correlation of the measurable and the non-measurable frequency
components of the motion. The parameters are chosen such that the
spectrum of the reconstructed force best approximates the measured
partial force spectrum. This approximative reconstruction requires
the use of numerical optimization techniques if the force model is
nonlinear in the parameters\cite{Forchheimer2012}. Analytic solutions
can be obtained if the model is linear in the parameters\cite{Hutter2010a,Raman2011,Platz2012a}.
Such a linear model of particular interest is the polynomial, as it
constitutes a general expansion of the tip-surface force.

Polynomial force reconstruction methods have been proposed theoretically
and tested on simulated data for intermodulation AFM (ImAFM)\cite{Hutter2010a,Platz2012a}.
Here, we demonstrate, for the first time polynomial force reconstruction
on experimental ImAFM data and compare it with reconstruction based
on amplitude-dependence force spectroscopy (ADFS)\cite{Platz2013}.
Moreover, fitting a force model to the polynomial reconstruction allows
for the extraction of properties like surface adhesion, sample stiffness
or interaction geometry. We demonstrate this extraction of surface
properties with high resolution stiffness maps on a blend of polystyrene
(PS) and poly(methyl methacrylate) (PMMA).

Polynomial reconstruction, and most other multifrequency methods,
work directly on the measured spectral data of the tip motion. Since
the tip motion can be very complicated, the interpretation of spectral
data often becomes rather difficult and alternative data representation
schemes might provide a better understanding of the tip-surface force.
Recently, we have shown how a narrow-band ImAFM measurement yields
the oscillation amplitude dependence of a force component $F_{I}$
in-phase with the sinusoidal tip motion and a force component $F_{Q}$
quadrature, or 90 degrees phase-shifted, to the tip motion\cite{Platz2012b}.
Here we show how polynomial force reconstruction can be perfomed within
the context of this picture of two force quadratures.

\section{Results and Discussion}

\subsection{Polynomial force reconstruction from spectral data}

In narrow band AFM the tip dynamics as a function of time $z(t)$
is usually described by a harmonic oscillator\cite{Rodrguez2002,Melcher2007},
subject to an external drive force and a time-dependent tip-surface
force
\begin{equation}
\ddot{z}+\frac{\omega_{0}}{Q}\dot{z}+k_{\mathrm{c}}(z-h)=F_{\mathrm{drive}}(t)+F_{\mathrm{ts}}(t)\label{eq:em-ho}
\end{equation}
where the dot denotes differentiation with respect to time, $\omega_{0}$,
$Q$ and $k_{\mathrm{c}}$ are the mode's resonance frequency, quality
factor and spring constant respectively, and $h$ is the static equilibrium
position of the tip above the surface. One should note that the time-dependence
of the tip-surface force $F_{ts}$ can be considered as an implicit
time-dependence, since it is assumed that the tip-surface interaction
depends on the instantaneous tip position $z$ and velocity $\dot{z}$
which are functions of time. In Fourier space equation (\ref{eq:em-ho})
becomes
\begin{equation}
\hat{z}(\omega)=\hat{\chi}(\omega)\left(\hat{F}_{\mathrm{drive}}(\omega)+\hat{F}_{\mathrm{ts}}(\omega)\right)\label{eq:em-fourier}
\end{equation}
where the linear response function 
\begin{equation}
\hat{\chi}(\omega)=\frac{\nicefrac{\omega_{0}^{2}}{k}}{\omega_{0}^{2}-\omega+i\frac{\omega_{0}\omega}{Q}}\label{eq:response-func}
\end{equation}
determines the tip response to a sinusoidal force applied at the frequency
$\omega$. The drive force can readily be determined from a measurement
of the tip motion far away from the surface, $z_{\mathrm{free}}(t)$,
where the tip-surface force is zero,
\begin{equation}
\hat{F}_{\mathrm{drive}}(\omega)=\hat{\chi}^{-1}(\omega)\hat{z}_{\mathrm{free}}(\omega)
\end{equation}
If the broad-band tip response close to the surface $\hat{z}(\omega)$
is known, one can easily solve equation (\ref{eq:em-fourier}) for
the spectrum of the tip surface force
\begin{equation}
\hat{F}_{\mathrm{ts}}(\omega)=\hat{\chi}^{-1}(\omega)\left(\hat{z}(\omega)-\hat{z}_{\mathrm{free}}(\omega)\right)\label{eq:force-fourier}
\end{equation}
With the inverse Fourier transform, the time-dependent force acting
on the tip can be readily determined from equation (\ref{eq:force-fourier}).

Since the result of an experiment is a vector $\underline{z}$ of
time-discrete samples of the continuous signal $z(t)$ during a time
window of length $T=1/\Delta\omega$, the Fourier transform can be
expressed using a unitary matrix $\underline{\underline{\mathcal{F}}}$
, 
\begin{eqnarray}
\hat{\underline{z}} & = & \mathcal{\underline{\underline{F}}\ }\underline{z}\\
\underline{z} & = & \mathcal{\underline{\underline{\hat{F}}}}^{-1}\ \underline{\hat{z}}
\end{eqnarray}
where a single underline denotes a vector and a double underline a
matrix. In a real experiment only a partial motion spectrum $\underline{\hat{z}}_{\mathrm{m}}$
can be measured since the cantilever's linear transfer function $\hat{\chi}$
suppresses response far away from resonance . Mathematically, this
can be expressed with a diagonal windowing matrix operator $\underline{\underline{\hat{W}}}$
that sets all frequency components outside the resonant detection
band to zero. The measured spectrum is then given by 
\begin{equation}
\underline{\hat{z}}_{\mathrm{m}}=\underline{\underline{\hat{W}}}\ \underline{\hat{z}}
\end{equation}
and equation (\ref{eq:force-fourier}) for the measured data becomes
\begin{equation}
\underline{\underline{\hat{\chi}}}^{-1}\left(\underline{\hat{z}}_{\mathrm{m}}-\underline{\hat{z}}_{\mathrm{free,m}}\right)=\underline{\underline{\hat{W}}}\ \underline{\underline{\hat{\chi}}}^{-1}\left(\underline{\hat{z}}-\underline{\hat{z}}_{\mathrm{free}}\right)=\underline{\underline{\hat{W}}}\ \underline{\hat{F}}_{\mathrm{ts}}\label{eq:matrix-force-fourier}
\end{equation}
Since $\hat{W}$ is not invertable we cannot determine the complete
force spectrum $\hat{F}_{\mathrm{ts}}$ from equation (\ref{eq:matrix-force-fourier})
and thus the time-dependent force remains unknown. To reconstruct
the complete force spectrum from the measured partial motion spectrum
$\hat{z}_{\mathrm{m}}$ we expand the tip-surface force into a finite
series from a set of basis functions
\begin{equation}
F_{\mathrm{ts}}(z,\dot{z})=\sum_{n=0}^{N}f_{n}(z,\dot{z})g_{n}
\end{equation}
Our assumption that the force can be expanded in this manner results
in a correlation of the unknown frequency components of the force
with the measurable components. A common choice for the functions
$f_{n}$ to model conservative forces are monomials\cite{Hutter2010a}
\begin{equation}
f_{n}(z,\dot{z})=(z-h)^{n}\label{eq:monomials}
\end{equation}
but also other basis functions like 
\begin{equation}
f_{m}(z,\dot{z})=\begin{cases}
(z-h)^{m/2} & m\ \mathrm{even}\\
\dot{z}(z-h)^{(m+1)/2} & m\ \mathrm{odd}
\end{cases}
\end{equation}
for the representation of position-dependent viscosites have been
considered\cite{Platz2012a}. For a measured tip motion the force
vector $\underline{F}_{ts}$ can the be approximated as 
\begin{equation}
\underline{F}_{\mathrm{ts}}\simeq\underline{\underline{H}}\ \underline{g}\label{eq:force-h-coord-space}
\end{equation}
where the coupling matrix $H$ is given by
\begin{equation}
\underline{\underline{H}}=\underline{\underline{H}}(\underline{z}_{\mathrm{m}},\underline{\dot{z}}_{\mathrm{m}})=\left(\begin{array}{cccc}
\vdots & \vdots &  & \vdots\\
\underline{f}_{0}(\underline{z}_{\mathrm{m}},\underline{\dot{z}}_{\mathrm{m}}) & \underline{f}_{1}(\underline{z}_{\mathrm{m}},\underline{\dot{z}}_{\mathrm{m}}) & \cdots & \underline{f}_{N}(\underline{z}_{\mathrm{m}},\underline{\dot{z}}_{\mathrm{m}})\\
\vdots & \vdots &  & \vdots
\end{array}\right)\label{eq:h-matrix-coord-space}
\end{equation}
in which the columns are formed by the vectors $\underline{f}_{0},\underline{f}_{1},\ldots,\underline{f}_{n}$
evaluated at the measured discrete tip positions and velocities. Here,
we assume that the measured, or windowed tip motion $\underline{z}_{m}$
is a good approximation of the complete tip motion $\underline{z}$.
In Fourier space equation (\ref{eq:force-h-coord-space}) becomes
\begin{equation}
\underline{\hat{F}}=\mathcal{\underline{\underline{F}}}\ \underline{\underline{H}}\ \underline{g}\equiv\underline{\underline{\hat{H}}}\ \underline{g}
\end{equation}
where
\begin{equation}
\underline{\underline{\hat{H}}}=\underline{\underline{\hat{H}}}(z_{\mathrm{m}},\dot{z}_{\mathrm{m}})=\left(\begin{array}{cccc}
\vdots & \vdots &  & \vdots\\
\mathcal{\underline{\underline{F}}}\ \underline{f}_{0}(\underline{z}_{\mathrm{m}},\underline{\dot{z}}_{\mathrm{m}}) & \mathcal{\underline{\underline{F}}}\ \underline{f}_{1}(\underline{z}_{\mathrm{m}},\underline{\dot{z}}_{\mathrm{m}}) & \cdots\mathcal{\underline{\underline{F}}}\  & \underline{f}_{N}(\underline{z}_{\mathrm{m}},\underline{\dot{z}}_{\mathrm{m}})\\
\vdots & \vdots &  & \vdots
\end{array}\right)\label{eq:h.matrix-fourier}
\end{equation}
The force matrix equation (\ref{eq:matrix-force-fourier}) can then
be written as 
\begin{equation}
\hat{\underline{\underline{W}}}\ \underline{\underline{\hat{H}}}\ \underline{g}=\underline{\hat{\chi}}^{-1}\left(\underline{\hat{z}}_{\mathrm{m}}-\underline{\hat{z}}_{\mathrm{free,m}}\right).\label{eq:matrix-window-force-fourier}
\end{equation}
We introduce 
\begin{equation}
\underline{\underline{\hat{H}}}_{\mathrm{m}}=\underline{\underline{\hat{W}}}\ \underline{\underline{\hat{H}}}
\end{equation}
and solve equation (\ref{eq:matrix-window-force-fourier}) for $\underline{g}$
such that 
\begin{equation}
\underline{g}=\underline{\underline{\hat{H}}}_{\mathrm{m}}^{+}\ \underline{\underline{\hat{\chi}}}^{-1}\left(\underline{\hat{z}}_{\mathrm{m}}-\underline{\hat{z}}_{\mathrm{free,m}}\right)\label{eq:coef-solution}
\end{equation}
where $\underline{\underline{\hat{H}}}_{\mathrm{m}}^{+}$ denotes
the pseudo-inverse of $\underline{\underline{\hat{H}}}_{\mathrm{m}}$.
If there exists an unique solution for the coefficient vector $\underline{g}$,
the matrix $\underline{\underline{\hat{H}}}_{\mathrm{m}}^{+}$ equals
the exact inverse of $\underline{\underline{\hat{H}}}_{\mathrm{m}}$
. If there is more than one solution for $\underline{g}$, equation
(\ref{eq:coef-solution}) computes the solution for which the the
vector $\underline{g}$ has minimum length. If no solution to equation
(\ref{eq:matrix-window-force-fourier}) exists the pseudo-inverse
$\underline{\underline{\hat{H}}}_{\mathrm{m}}^{+}$ approximates the
inverse in a least-square sense.

The matrix $\underline{\underline{\hat{H}}}_{\mathrm{m}}$ can be
rapidly computed from equation (\ref{eq:h.matrix-fourier}) using
the Fast Fourier Transform (FFT) algorithm. Therefore, equation (\ref{eq:coef-solution})
provides a efficient way to determine the expansion coefficients $\underline{g}$
of the the tip-surface force. However, special care should be taken
to avoid aliasing effects due to the finite sampling of the data.
To increase the numerical stability of equation (\ref{eq:coef-solution})
it is advantageous to normalize $\underline{z}_{\mathrm{m}}$ and
$\underline{\dot{z}}_{\mathrm{m}}$ such that the largest absolute
value of any vector element is 1. This normalization can be interpreted
as a pre-conditioning procedure for the matrix $\underline{\underline{\hat{H}}}_{\mathrm{m}}$
.

\begin{figure}
\centering{}
\includegraphics{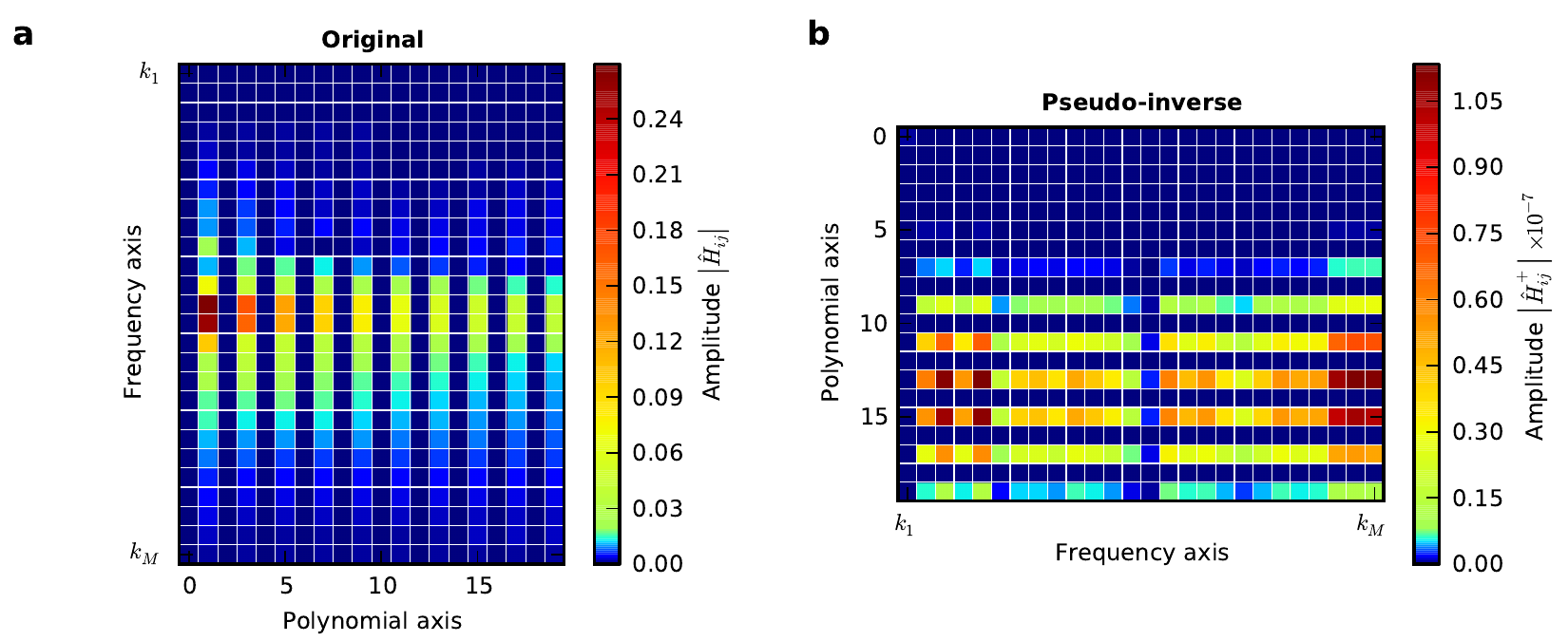}
\caption{Visualization of the $\underline{\underline{\hat{H}}}_{\mathrm{m}}$
matrix (a) and its pseudo-inverse (b). Only rows with non-zero elements
are displayed.\label{fig:h-matrix} }
\end{figure}

To further investigate which information about the tip-surface force
can be extracted, we focus on the monomial expansion basis defined
in equation (\ref{eq:monomials}) and the case of narrow band ImAFM
where the windowing matrix is given by
\begin{equation}
\hat{W}_{ij}=\begin{cases}
\delta_{ij} & k_{1}\leq i\leq k_{M}\\
0 & \mathrm{else}
\end{cases}
\end{equation}
with $\delta_{ij}$ being the Kronecker delta, $k_{1}\Delta\omega$
the lower frequency limit of the resonant detection band and $k_{M}\Delta\omega$
the upper limit. In figure \ref{fig:h-matrix} we plot the absolute
values of the components of the matrix $\underline{\underline{\hat{H}}}_{\mathrm{m}}$
for experimental data. One could imagine applying different windowing
matrices when building $\underline{\underline{\hat{H}}}_{\mathrm{m}}$,
for example one which is weighted by the signal-to-noise ratio at
each frequency. 
The absolute value of the element of $\underline{\underline{\hat{H}}}_{\mathrm{m}}$with
the index $(i,j)$ is a measure of how much the j\textsuperscript{th}
element $g_{j}$ of the expansion coefficient vector $\underline{g}$
, contributes to the force at the i\textsuperscript{th} frequency
$i\Delta\omega$ in the force spectrum $\underline{\hat{F}}_{\mathrm{ts}}$.
It is apparent from figure \ref{fig:h-matrix} that only polynomial
coefficients of odd order contribute to the force measured in the
resonant detection band when two drives close to resonance are used
in ImAFM. Thus, equation (\ref{eq:coef-solution}) only yields the
odd coefficients in the polynomial force expansion and the resulting
polynomial force is odd with respect to $z=h$. To determine the missing
even coefficients we assume that the tip-surface force is zero for
$z\geq z_{\mathrm{non-interacting}}$. With this assumption we fit
the even polynomial coefficients while keeping the odd coefficients
constant. This reconstruction method has been extensively tested and
its accuracy verified with simulated data\cite{Platz2012a}. In the
following we will show results for experimental data.

\subsection{Polynomial force reconstruction during slow surface approach}

To demonstarte the capabilities of the polynomial force reconstruction
we perform an ImAFM approach measurement on a silicon oxide surface.
In this measurement two drive frequencies close to resonance result
in a beat-like tip motion, with rapid sinusoidal oscillations and
a slowly varying amplitude. The AFM z-piezo moves solowly towards
the surface, such that during one beat period the static tip height
above the surface can be considered to be constant. 

\begin{figure}
\centering{}
\includegraphics{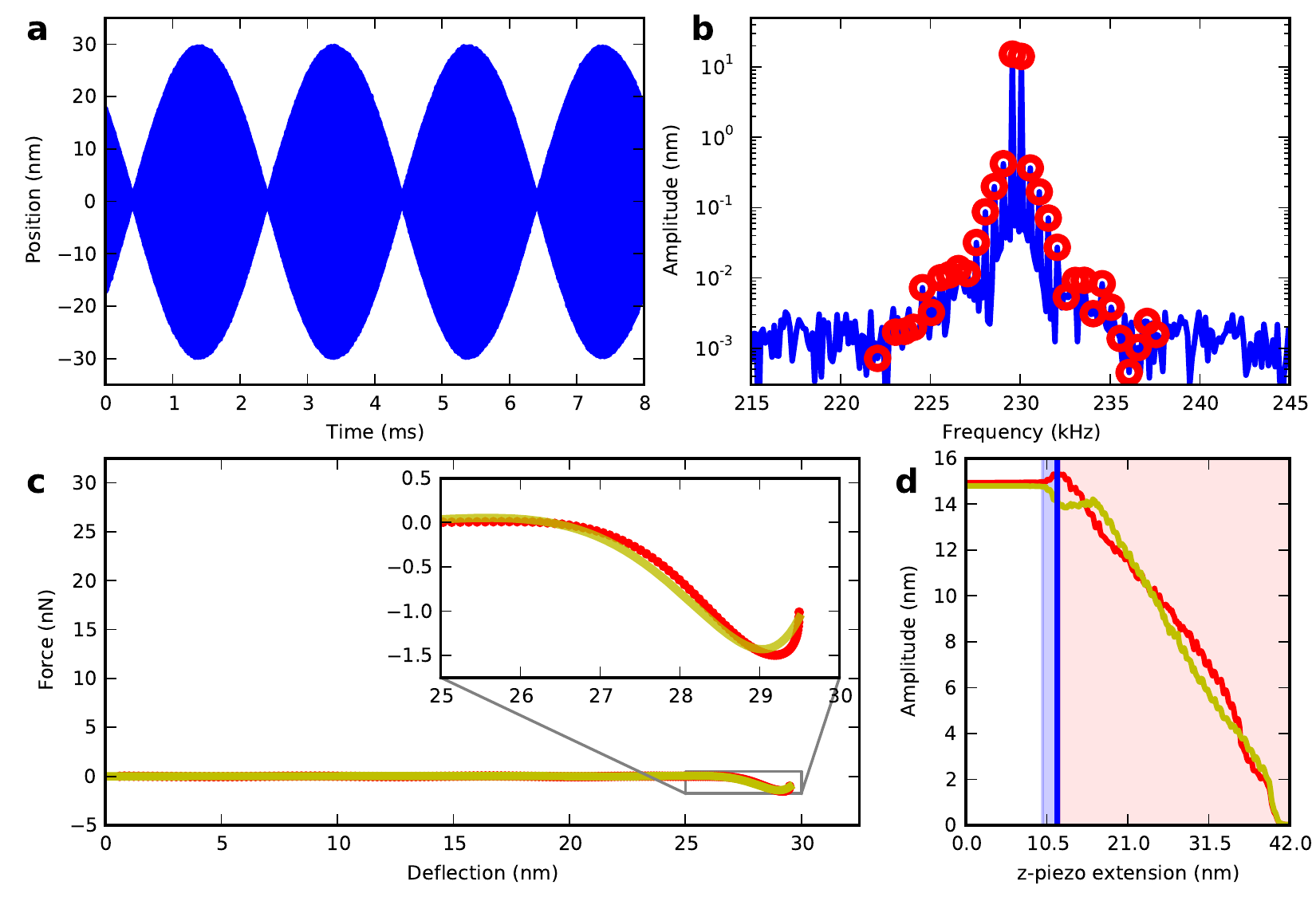}
\caption{One frame from a surface approach movie (link to movie) showing the
very onset of repulsive forces. The beating waveform (a) has the intermodulation
spectrum around resonance (b) where red circles are analyzed to reconstruct
the tip-surface force (c) using both both polynomial (yellow) and
ADFS (red) methods. The z-piezo extension $z_{\mathrm{piezo}}=11.8\ \mathrm{nm}$
is indicated by the blue vertical line in (d) which the displays the
amplitudes at the two drive frequencies. The interaction is purely
attractive in the blue shaded area of (d) , becoming repulsive in
the red shaded area.\label{fig:frame-attractive}}
\end{figure}

Figure \ref{fig:frame-attractive} shows one frame from a movie (supporting
information file 1) visualizing the measurement. For four consecutive
beats in the time domain (a) the corresponding amplitude spectrum
around the first resonance is displayed in (b) where the components
or partial spectrum used for force reconstruction are marked with
red circles. The polynomial force reconstruction is plotted (yellow
solid line) in figure \ref{fig:frame-attractive}c together with an
ADFS reconstruction using the same data (red circles). In figure \ref{fig:frame-attractive}d
the amplitudes of the tip motion at the lower (red) and the higher
drive frequency (yellow) are shown as functions of the z-piezo extension,
and the vertical blue line indicates the current z-piezo extension.

Far away from the surface the tip does not experience any surface
force and the motion spectrum exhibits response only at the driven
frequencies (supporting information file 1). Consequently, the reconstructed
force is zero. As the surface is approached the attractive force regime
due to the van-der-Waals forces between the tip and the surface is
reached. In this regime new frequency components appear in the motion
spectrum, so-called intermodulation products. Note that in the time
domain the distortion of the signal is barely visible. Both polynomial
and ADFS reconstruction show a growing attractive interaction until
a force minimum of -1.75 nN is reached at a piezo extension of 11.8
nm. At this point the tip experiences hard mechanical impacts on the
sample surface near the beat maximum which are manifest in the sharp
onset of repulsive force in the polynomial and ADFS reconstruction
(figure \ref{fig:frame-attractive}c). 

\begin{figure}
\begin{centering}
\includegraphics{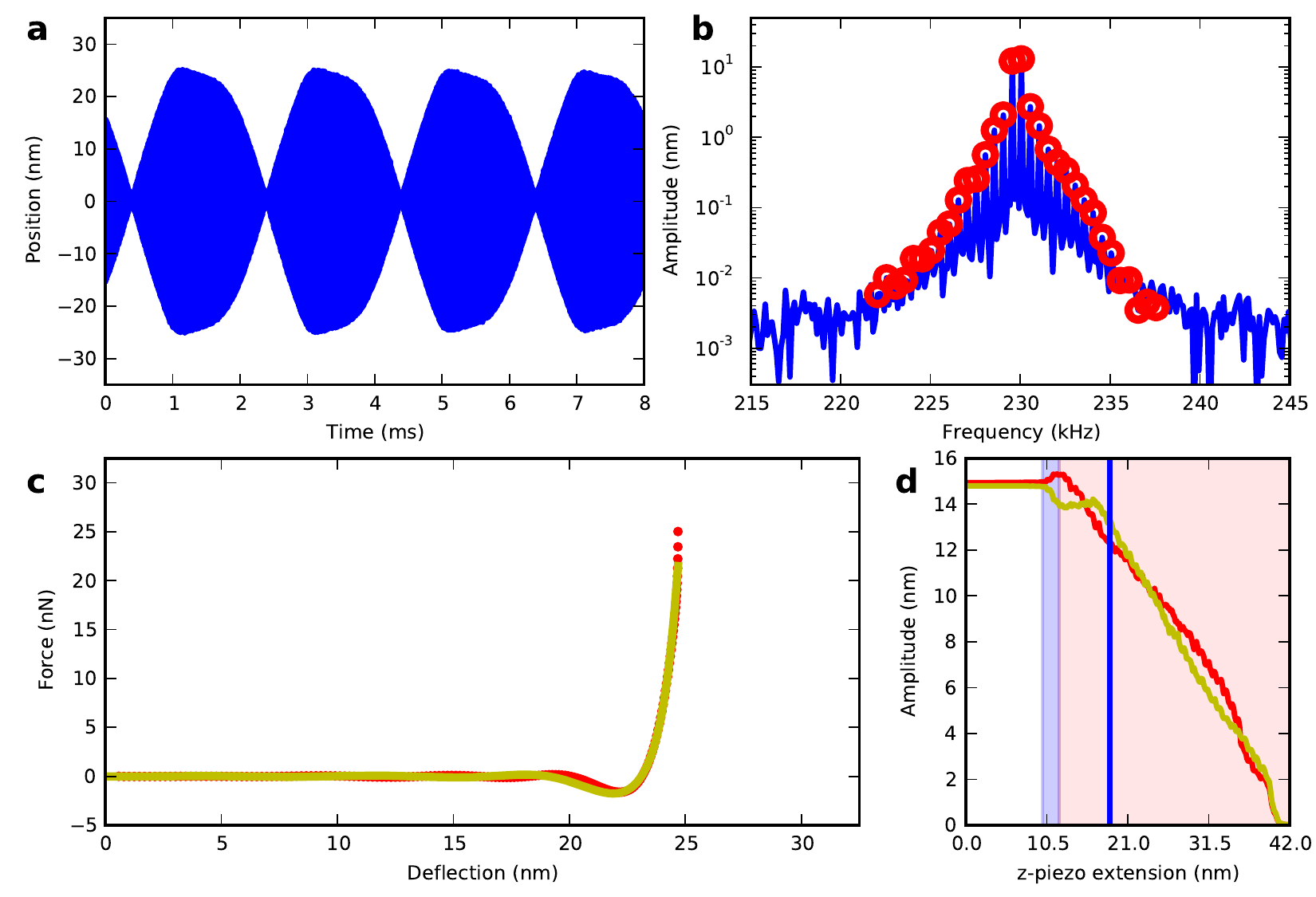}
\par\end{centering}
\caption{One frame from a surface approach movie (link to movie) showing strongly
repulsive forces. The beating waveform (a) has the intermodulation
spectrum around resonance (b) where red circles are analyzed to reconstruct
the tip-surface force (c) using both both polynomial (yellow) and
ADFS (red) methods. The z-piezo extension $z_{\mathrm{piezo}}=18.7\ \mathrm{nm}$
is indicated by the blue vertical line in (d) which the displays the
amplitudes at the two drive frequencies. The interaction is purely
attractive in the blue shaded area of (d) , becoming repulsive in
the red shaded area.\label{fig:frame-repulsive}}
\end{figure}

As the z-piezo further extends the tip indents deeper into the surface
and experiences stronger repulsive forces as shown in figure \ref{fig:frame-repulsive}
where one frame of the movie at a piezo extension of 18.7 nm is shown.
In this repulsive regime the polynomial and the ADFS reconstruction
agree very well. However, the force minimum is slightly sharper shape
with the ADFS reconstruction, which is not constrained to be continuous
in high-order derivatives, as for a polynomial. The repulsive force
reaches its maximum of 31 nN at a z-piezo extension of 23.6. Moving
closer to the surface the maximum force during one beat decreases
until the oscillation vanishes. 

During the whole surface approach the polynomial and the ADFS force
reconstruction agree very well, indicating that the polynomial reconstruction
accurately reproduces the force. The shape of the reconstructed force
is very stable during the entire approach for both polynomial and
ADFS reconstruction which is a result of the high signal-to-noise
ratio for the measured frequency components close to resonance. The
stability of the reconstruction during approach gives us confidence
in the method's ability to accurately reconstruct sharp features in
the force curve, like the force minimum.

\subsection{Surface parameter mapping}

Material scientists are often interested in determining surface properties
with high spatial resolution. ImAFM measurements can be preformed
while scanning a sample surface\cite{Platz2008,Platz2010}, enabling
polynomial force reconstruction in every point of an AFM image. A
specific force model can then be fitted to the complete force curve
or parts thereof, generating a map of the model parameters over the
complete surface. To demonstrate this we scanned the surface of a
PS/PMMA blend with ImAFM. To the repulsive part of the polynomial
force reconstruction we fit a Derjaguin-Muller-Toropov (DMT) force
model\cite{DERJAGUIN1975} of the form
\[
F_{\mathrm{rep}}(z)=F_{\mathrm{min}}+\epsilon(z-z_{\mathrm{min}})^{3/2}
\]
where $z_{\mathrm{min}}$ is the position of the force minimum $F_{\mathrm{min}}$
and $\epsilon$ is the DMT stiffness factor which depends on the tip
radius and the effective stiffness of the tip-surface system. On should
be aware of the fact that macroscopic force models like the DMT model
might not be applicable on the nanoscale\cite{Luan2005} and that
tip shape and surface topography lead to an interaction geometry which
is different from the model geometry of a perfect sphere and a perfect
flat\cite{Platz2013}. Moreover, the DMT model does not account for
adhesive forces in the contact regime which we try to circumvent by
using tips with small radii. While the DMT model provides sufficient
insight into material properties, the extracted numerical values of
the DMT parameters should not be expected to agree with values for
the bulk material.

\begin{figure}
\begin{centering}
\includegraphics{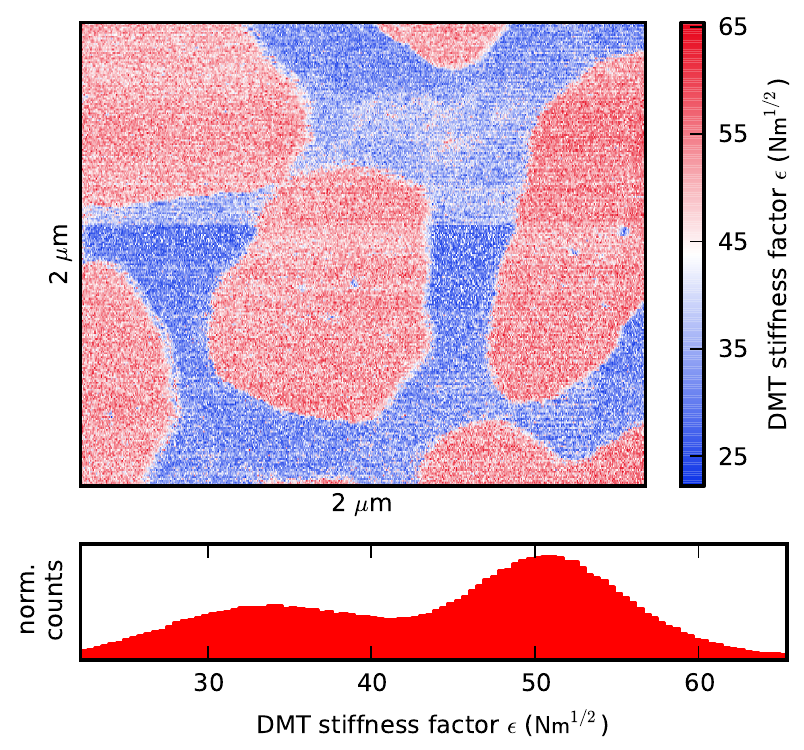}
\caption{A surface property map showing the DMT stiffness factor $\epsilon$
{[}$\mathrm{GPa\ m^{1/2}}$ {]} with a histogram of the stiffness
factor.\label{fig:parameter-map} }

\par\end{centering}

\end{figure}

In figure \ref{fig:parameter-map} a map of the DMT stiffness factor
is shown. Even thought, the two polymers are very similar in stiffness
at room temperatures\cite{Brandrup2005}, two domains of different
stiffness are clearly visible in the stiffness factor map. The stiffer
domains are PMMA-rich and 10 nm higher than the surrounding matrix
which is PS-rich and is a factor of two softer than the PMMA-domains.
Similar results on the same model polymer system have been obtained
with higher harmonics methods\cite{Sahin2007,Sahin2008a} and ADFS\cite{Platz2013}.

\subsection{Polynomial reconstruction from force quadrature data}

Polynomial force reconstruction is an approximative reconstruction
method applied to spectral data obtained from Fourier analysis of
the tip motion. The general idea, to determine the parameters of a
force model such that an experimental observable is best approximated,
is not limited to spectral data. An alternative to the spectral representation
is a picture which which represents the data in terms of two force
quadratures. The force quadrature $F_{I}$ is the component of the
(time-dependent) tip-surface force that is in phase with the sinusoidal
tip motion, the quadrature $F_{Q}$ is the force component which is
phase-shifted 90 degrees to the tip motion\cite{Platz2012d}.
\begin{eqnarray}
F_{I}(A,\bar{\omega},h) & = & \frac{1}{T}\int_{0}^{T}F_{\mathrm{ts}}\left(A\cos(\bar{\omega}t)+h,-\bar{\omega}A\sin(\bar{\omega}t)\right)\cos(\bar{\omega}t)dt\label{eq:fi-int}\\
F_{Q}(A,\bar{\omega},h) & = & \frac{1}{T}\int_{0}^{T}F_{\mathrm{ts}}\left(A\cos(\bar{\omega}t)+h,-\omega A\sin(\bar{\omega}t)\right)\sin(\bar{\omega}t)dt\label{eq:fq-int}
\end{eqnarray}
where the force quadratures are functions of the oscillation amplitude
$A$, the oscillation frequency $\bar{\omega}$ and the static tip
height $h$, all of which are constant during each oscillation cycle.
However, here we consider only the amplitude dependence of $F_{I}$
and $F_{Q}$ which can be rapidly measured with ImAFM using a single
oscillation cycle analysis that is based on a separation of time scales\cite{Platz2012b}. 

The representation of the measurement result in terms of the force
quadratures $F_{I}$ and $F_{Q}$ has the advantage that they are
directly connected to the tip-surface force and independent of the
actual complicated multifrequency tip motion. With spectral data certain
points on the tip-surface force curve will receive greater weight
if the tip spends more time at these positions. On the $F_{I}(A)$
and $F_{Q}(A)$ curves the weight at each amplitude can be controlled
by design. Furthermore, distortions due to feedback artifacts can
easily be removed from the $F_{I}(A)$ and $F_{Q}(A)$ curves and
both conservative and dissipative forces can be analyzed separately. 

To demonstrate approximative force reconstruction on force quadrature
data, we consider again a conservative polynomial force representation
as in equation (\ref{eq:monomials}). For such a force $F_{Q}(A)=0$
and equation (\ref{eq:fi-int}) becomes
\begin{equation}
F_{I}(A)=\frac{1}{T}\sum_{n=0}^{N}g_{n}A^{n}\int_{0}^{T}\cos^{n+1}(\omega t)dt\label{eq:fi-poly}
\end{equation}
The integral is non-zero only for odd $n$ and by using 
\[
\cos^{n}(\theta)=\frac{1}{2^{n}}\left(\begin{array}{c}
n\\
n/2
\end{array}\right)+\frac{1}{2^{n-1}}\sum_{k=0}^{\frac{n}{2}-1}\left(\begin{array}{c}
n\\
k
\end{array}\right)\cos\left((n-2k)\theta\right)
\]
where $\left(\begin{array}{c}
n\\
k
\end{array}\right)$ is the binomial coefficient, equation (\ref{eq:fi-poly}) becomes
\[
F_{I}(A)=\sum_{\begin{array}{c}
n=0\\
n\ \mathrm{odd}
\end{array}}^{N}\frac{1}{2^{n}}\left(\begin{array}{c}
n+1\\
\frac{n+1}{2}
\end{array}\right)g_{n}A^{n}\equiv\sum_{\begin{array}{c}
n=0\\
n\ \mathrm{odd}
\end{array}}^{N}\tilde{g}_{n}A^{n}
\]
which implies that the odd polynomial coefficients of the force expansion
can be obtained by simply rescaling of the coefficients of a polynomial
approximation of the $F_{I}(A)$ curve,
\[
g_{n}=\left[\frac{1}{2^{n+1}}\left(\begin{array}{c}
n+1\\
\frac{n+1}{2}
\end{array}\right)\right]^{-1}\tilde{g}_{n}
\]
To obtain the even coefficients we apply the same algorithm as for
reconstruction from spectral data.

We obtain a polynomial approximation of $F_{I}(A)$ by simple polynomial
fit to equidistant discrete points on the $F_{I}(A)$ curve. Alternative
methods like expansion in orthogonal polynomials or different types
of interpolation polynomials with different convergence properties
can also be applied. The polynomial reconstruction based on force
quadrature data can be implemented even more efficiently than the
reconstruction on spectral data since multiple Fourier transforms
to construct the coupling matrix $\underline{\underline{\hat{H}}}_{\mathrm{m}}$
in equation (\ref{eq:h.matrix-fourier}) are not required. 

\begin{figure}

\begin{centering}
\includegraphics{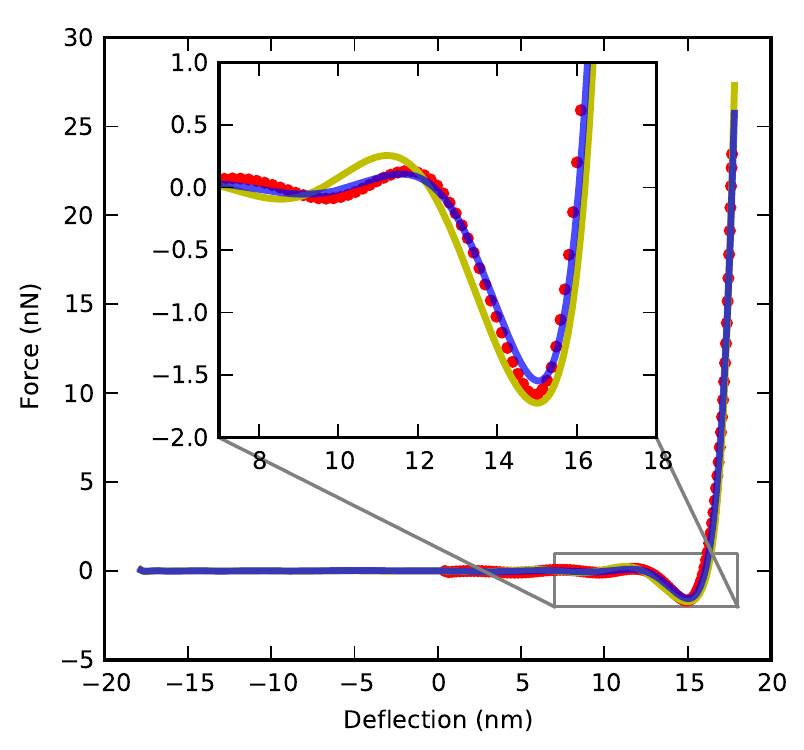}
\caption{Comparison of force reconstructions for a tip interacting with silicon
oxide. Polynomial reconstruction from spectral data (yellow) and force
quadrature data (blue) and ADFS reconstruction (red circles) are in
excellent agreement.\label{fig:poly-quadrature-reconstruction}}

\par\end{centering}

\end{figure}

To validate the equivalence of polynomial force reconstruction on
spectral and force quadrature data, we consider the ImAFM approach
measurement on silicon oxide described above. From the tip motion
at a z-piezo extension of 25.6 nm we compute the $F_{I}(A)$ curve
and remove all data points for which the oscillation amplitude was
decreasing. From the polynomial approximation of the $F_{I}(A)$ curve
we obtain the force polynomial as described above. The resulting reconstruction
is shown in figure \ref{fig:poly-quadrature-reconstruction} (blue
line) together with the reconstruction from spectral data (yellow
line) and an ADFS reconstruction (red circles). Over the full range
of oscillation both curves agree extremely well. Excellent agreement
is also observed at all other z-piezo extensions.

\section{Conclusions}

Multifrequency AFM opens the window to a wide variety of novel AFM
measurement techniques enabling a much improved understanding of the
forces between the tip and the surface. We showed that polynomial
force reconstruction is an intuitive and powerful method to approximate
this interaction, we demonstrated the method's use for accurate and
detailed force measurement and for high resolution surface parameter
mapping with experimental data.

As the field of multifrequency AFM continues to evolve, new alternative
data representation schemes can help to simplify analysis and extract
more specific property of the tip-surface interaction. The force quadrature
picture is such a scheme which decouples information about the tip-surface
interaction from the actual tip motion. We showed how the general
idea behind approximative force reconstruction can be adapted to the
force quadrature picture and we introduced an efficient way to extract
the polynomial coefficients from the force quadratures.

We hope that in the future polynomial force reconstruction will be
a useful method for many scientists and that new data representation
schemes will inspire innovative analysis methods.

\section{Experimental}

The silicon oxide sample was cleaned in an oxygen plasma before performing
measurements in a Bruker Dimension 3100 AFM system. The cantilever
(Bruker MPP-11120) was calibrated using a non-invasive thermal method\cite{Higgins2006}
and had a resonance frequency of $f_{0}=229.802\ \mathrm{kHz}$, a
quality factor of $Q=396.9$ and a spring constant of $k_{\mathrm{c}}=16.0\ \mathrm{nm^{-1}}$.
The slow surface approach velocity was $2\ \mathrm{nm\, s^{-1}}$.

PS ($M_{w}=280\ \mathrm{kDa}$, Sigma-Aldrich) and PMMA ($M_{w}=120\ \mathrm{kDa}$,
Sigma-Aldrich) were spin-cast from toluene solution with a concentration
of 0.53 \%wt at ratio of 3:1 (PMMA:PS). The sample was scanned in
a Bruker Multimode 2 AFM system with a cantilever BS 300Al-G (Budget
Sensors) having a resonance frequency $f_{0}=343.379\ \mathrm{kHz}$,
quality factor $Q=556.9$ and spring constant $k_{\mathrm{c}}=35.1\ \mathrm{nm^{-1}}$.
The maximum free oscillation amplitude close to the surfacce was 30
nm and we scanned an image with $256\times1024$ pixel within 17 minutes.

For all measurements we used a intermodulation lockin analyzer (IMP
2-32, Intermodulation Products AB) which synchronizes the signal generation
and acquisition for measurement of the multifrequency response\cite{Tholen2011}.

\section{Supporting information}

Movie showing the tip motion and the reconstructed tip-surface during
a ImAFM approach measurement
\begin{itemize}
\item Supporting information file 1

\begin{itemize}
\item File name: s1.mp4
\item File format: mp4
\item Title: Tip-surface force during ImAFM approach measurement
\end{itemize}
\end{itemize}

\section{Acknowledgments}
The authors acknowledge financial support from the Knut and Alice
Wallenberg Foundation, the Swedish Research Council (VR), the Swedish
Government Agency for Innovation Systems (VINNOVA) and the Olle Engkvist
Foundation.


\begin{thebibliography}{32}
\expandafter\ifx\csname natexlab\endcsname\relax\def\natexlab#1{#1}\fi
\expandafter\ifx\csname bibnamefont\endcsname\relax
  \def\bibnamefont#1{#1}\fi
\expandafter\ifx\csname bibfnamefont\endcsname\relax
  \def\bibfnamefont#1{#1}\fi
\expandafter\ifx\csname citenamefont\endcsname\relax
  \def\citenamefont#1{#1}\fi
\expandafter\ifx\csname url\endcsname\relax
  \def\url#1{\texttt{#1}}\fi
\expandafter\ifx\csname urlprefix\endcsname\relax\def\urlprefix{URL }\fi
\providecommand{\bibinfo}[2]{#2}
\providecommand{\eprint}[2][]{\url{#2}}

\bibitem[{\citenamefont{Ohnesorge and Binnig}(1993)}]{Ohnesorge1993}
\bibinfo{author}{\bibfnamefont{F.}~\bibnamefont{Ohnesorge}} \bibnamefont{and}
  \bibinfo{author}{\bibfnamefont{G.}~\bibnamefont{Binnig}},
  \bibinfo{journal}{Science} \textbf{\bibinfo{volume}{260}},
  \bibinfo{pages}{1451} (\bibinfo{year}{1993}), ISSN \bibinfo{issn}{0036-8075}.

\bibitem[{\citenamefont{Giessibl}(1995)}]{Giessibl1995}
\bibinfo{author}{\bibfnamefont{F.~J.} \bibnamefont{Giessibl}},
  \bibinfo{journal}{Science} \textbf{\bibinfo{volume}{267}},
  \bibinfo{pages}{68} (\bibinfo{year}{1995}), ISSN \bibinfo{issn}{0036-8075}.

\bibitem[{\citenamefont{Fukuma et~al.}(2005)\citenamefont{Fukuma, Kobayashi,
  Matsushige, and Yamada}}]{Fukuma2005}
\bibinfo{author}{\bibfnamefont{T.}~\bibnamefont{Fukuma}},
  \bibinfo{author}{\bibfnamefont{K.}~\bibnamefont{Kobayashi}},
  \bibinfo{author}{\bibfnamefont{K.}~\bibnamefont{Matsushige}},
  \bibnamefont{and} \bibinfo{author}{\bibfnamefont{H.}~\bibnamefont{Yamada}},
  \bibinfo{journal}{Applied Physics Letters} \textbf{\bibinfo{volume}{86}},
  \bibinfo{pages}{193108} (\bibinfo{year}{2005}), ISSN
  \bibinfo{issn}{00036951}.

\bibitem[{\citenamefont{Gan}(2009)}]{Gan2009}
\bibinfo{author}{\bibfnamefont{Y.}~\bibnamefont{Gan}},
  \bibinfo{journal}{Surface Science Reports} \textbf{\bibinfo{volume}{64}},
  \bibinfo{pages}{99} (\bibinfo{year}{2009}), ISSN \bibinfo{issn}{01675729}.

\bibitem[{\citenamefont{D\"{u}rig}(2000{\natexlab{a}})}]{Durig2000}
\bibinfo{author}{\bibfnamefont{U.}~\bibnamefont{D\"{u}rig}},
  \bibinfo{journal}{Applied Physics Letters} \textbf{\bibinfo{volume}{76}},
  \bibinfo{pages}{1203} (\bibinfo{year}{2000}{\natexlab{a}}), ISSN
  \bibinfo{issn}{00036951}.

\bibitem[{\citenamefont{Sader et~al.}(2005)\citenamefont{Sader, Uchihashi,
  Higgins, Farrell, Nakayama, and Jarvis}}]{Sader2005}
\bibinfo{author}{\bibfnamefont{J.~E.} \bibnamefont{Sader}},
  \bibinfo{author}{\bibfnamefont{T.}~\bibnamefont{Uchihashi}},
  \bibinfo{author}{\bibfnamefont{M.~J.} \bibnamefont{Higgins}},
  \bibinfo{author}{\bibfnamefont{A.}~\bibnamefont{Farrell}},
  \bibinfo{author}{\bibfnamefont{Y.}~\bibnamefont{Nakayama}}, \bibnamefont{and}
  \bibinfo{author}{\bibfnamefont{S.~P.} \bibnamefont{Jarvis}},
  \bibinfo{journal}{Nanotechnology} \textbf{\bibinfo{volume}{16}},
  \bibinfo{pages}{S94} (\bibinfo{year}{2005}), ISSN \bibinfo{issn}{0957-4484}.

\bibitem[{\citenamefont{H\"{o}lscher}(2006)}]{Holscher2006}
\bibinfo{author}{\bibfnamefont{H.}~\bibnamefont{H\"{o}lscher}},
  \bibinfo{journal}{Applied Physics Letters} \textbf{\bibinfo{volume}{89}},
  \bibinfo{pages}{123109} (\bibinfo{year}{2006}), ISSN
  \bibinfo{issn}{00036951}.

\bibitem[{\citenamefont{Lee and Jhe}(2006)}]{Lee2006}
\bibinfo{author}{\bibfnamefont{M.}~\bibnamefont{Lee}} \bibnamefont{and}
  \bibinfo{author}{\bibfnamefont{W.}~\bibnamefont{Jhe}},
  \bibinfo{journal}{Physical Review Letters} \textbf{\bibinfo{volume}{97}},
  \bibinfo{pages}{036104} (\bibinfo{year}{2006}), ISSN
  \bibinfo{issn}{0031-9007}.

\bibitem[{\citenamefont{Hu and Raman}(2008)}]{Hu2008}
\bibinfo{author}{\bibfnamefont{S.}~\bibnamefont{Hu}} \bibnamefont{and}
  \bibinfo{author}{\bibfnamefont{A.}~\bibnamefont{Raman}},
  \bibinfo{journal}{Nanotechnology} \textbf{\bibinfo{volume}{19}},
  \bibinfo{pages}{375704} (\bibinfo{year}{2008}), ISSN
  \bibinfo{issn}{0957-4484}.

\bibitem[{\citenamefont{Katan et~al.}(2009)\citenamefont{Katan, van Es, and
  Oosterkamp}}]{Katan2009}
\bibinfo{author}{\bibfnamefont{A.~J.} \bibnamefont{Katan}},
  \bibinfo{author}{\bibfnamefont{M.~H.} \bibnamefont{van Es}},
  \bibnamefont{and} \bibinfo{author}{\bibfnamefont{T.~H.}
  \bibnamefont{Oosterkamp}}, \bibinfo{journal}{Nanotechnology}
  \textbf{\bibinfo{volume}{20}}, \bibinfo{pages}{165703}
  (\bibinfo{year}{2009}), ISSN \bibinfo{issn}{1361-6528}.

\bibitem[{\citenamefont{Garc\'{\i}a and Herruzo}(2012)}]{Garcia2012}
\bibinfo{author}{\bibfnamefont{R.}~\bibnamefont{Garc\'{\i}a}} \bibnamefont{and}
  \bibinfo{author}{\bibfnamefont{E.~T.} \bibnamefont{Herruzo}},
  \bibinfo{journal}{Nature Nanotechnology} \textbf{\bibinfo{volume}{7}},
  \bibinfo{pages}{217} (\bibinfo{year}{2012}), ISSN \bibinfo{issn}{1748-3387}.

\bibitem[{\citenamefont{D\"{u}rig}(2000{\natexlab{b}})}]{Durig2000a}
\bibinfo{author}{\bibfnamefont{U.}~\bibnamefont{D\"{u}rig}},
  \bibinfo{journal}{New Journal of Physics} \textbf{\bibinfo{volume}{2}},
  \bibinfo{pages}{5} (\bibinfo{year}{2000}{\natexlab{b}}), ISSN
  \bibinfo{issn}{1367-2630}.

\bibitem[{\citenamefont{Stark et~al.}(2002)\citenamefont{Stark, Stark, Heckl,
  and Guckenberger}}]{Stark2002}
\bibinfo{author}{\bibfnamefont{M.}~\bibnamefont{Stark}},
  \bibinfo{author}{\bibfnamefont{R.~W.} \bibnamefont{Stark}},
  \bibinfo{author}{\bibfnamefont{W.~M.} \bibnamefont{Heckl}}, \bibnamefont{and}
  \bibinfo{author}{\bibfnamefont{R.}~\bibnamefont{Guckenberger}},
  \bibinfo{journal}{Proceedings of the National Academy of Sciences of the
  United States of America} \textbf{\bibinfo{volume}{99}},
  \bibinfo{pages}{8473} (\bibinfo{year}{2002}), ISSN \bibinfo{issn}{0027-8424}.

\bibitem[{\citenamefont{Legleiter et~al.}(2006)\citenamefont{Legleiter, Park,
  Cusick, and Kowalewski}}]{Legleiter2006}
\bibinfo{author}{\bibfnamefont{J.}~\bibnamefont{Legleiter}},
  \bibinfo{author}{\bibfnamefont{M.}~\bibnamefont{Park}},
  \bibinfo{author}{\bibfnamefont{B.}~\bibnamefont{Cusick}}, \bibnamefont{and}
  \bibinfo{author}{\bibfnamefont{T.}~\bibnamefont{Kowalewski}},
  \bibinfo{journal}{Proceedings of the National Academy of Sciences of the
  United States of America} \textbf{\bibinfo{volume}{103}},
  \bibinfo{pages}{4813} (\bibinfo{year}{2006}), ISSN \bibinfo{issn}{0027-8424}.

\bibitem[{\citenamefont{Sahin et~al.}(2007)\citenamefont{Sahin, Magonov, Su,
  Quate, and Solgaard}}]{Sahin2007}
\bibinfo{author}{\bibfnamefont{O.}~\bibnamefont{Sahin}},
  \bibinfo{author}{\bibfnamefont{S.}~\bibnamefont{Magonov}},
  \bibinfo{author}{\bibfnamefont{C.}~\bibnamefont{Su}},
  \bibinfo{author}{\bibfnamefont{C.~F.} \bibnamefont{Quate}}, \bibnamefont{and}
  \bibinfo{author}{\bibfnamefont{O.}~\bibnamefont{Solgaard}},
  \bibinfo{journal}{Nature nanotechnology} \textbf{\bibinfo{volume}{2}},
  \bibinfo{pages}{507} (\bibinfo{year}{2007}), ISSN \bibinfo{issn}{1748-3395}.

\bibitem[{\citenamefont{Forchheimer et~al.}(2012)\citenamefont{Forchheimer,
  Platz, Thol\'{e}n, and Haviland}}]{Forchheimer2012}
\bibinfo{author}{\bibfnamefont{D.}~\bibnamefont{Forchheimer}},
  \bibinfo{author}{\bibfnamefont{D.}~\bibnamefont{Platz}},
  \bibinfo{author}{\bibfnamefont{E.~A.} \bibnamefont{Thol\'{e}n}},
  \bibnamefont{and} \bibinfo{author}{\bibfnamefont{D.~B.}
  \bibnamefont{Haviland}}, \bibinfo{journal}{Physical Review B}
  \textbf{\bibinfo{volume}{85}}, \bibinfo{pages}{1} (\bibinfo{year}{2012}),
  ISSN \bibinfo{issn}{1098-0121}.

\bibitem[{\citenamefont{Hutter et~al.}(2010)\citenamefont{Hutter, Platz,
  Thol\'{e}n, Hansson, and Haviland}}]{Hutter2010a}
\bibinfo{author}{\bibfnamefont{C.}~\bibnamefont{Hutter}},
  \bibinfo{author}{\bibfnamefont{D.}~\bibnamefont{Platz}},
  \bibinfo{author}{\bibfnamefont{E.~A.} \bibnamefont{Thol\'{e}n}},
  \bibinfo{author}{\bibfnamefont{T.~H.} \bibnamefont{Hansson}},
  \bibnamefont{and} \bibinfo{author}{\bibfnamefont{D.~B.}
  \bibnamefont{Haviland}}, \bibinfo{journal}{Physical Review Letters}
  \textbf{\bibinfo{volume}{104}}, \bibinfo{pages}{1} (\bibinfo{year}{2010}),
  ISSN \bibinfo{issn}{0031-9007}.

\bibitem[{\citenamefont{Raman et~al.}(2011)\citenamefont{Raman, Trigueros,
  Cartagena, Stevenson, Susilo, Nauman, and Contera}}]{Raman2011}
\bibinfo{author}{\bibfnamefont{A.}~\bibnamefont{Raman}},
  \bibinfo{author}{\bibfnamefont{S.}~\bibnamefont{Trigueros}},
  \bibinfo{author}{\bibfnamefont{A.}~\bibnamefont{Cartagena}},
  \bibinfo{author}{\bibfnamefont{a.~P.~Z.} \bibnamefont{Stevenson}},
  \bibinfo{author}{\bibfnamefont{M.}~\bibnamefont{Susilo}},
  \bibinfo{author}{\bibfnamefont{E.}~\bibnamefont{Nauman}}, \bibnamefont{and}
  \bibinfo{author}{\bibfnamefont{S.~A.} \bibnamefont{Contera}},
  \bibinfo{journal}{Nature nanotechnology} \textbf{\bibinfo{volume}{6}},
  \bibinfo{pages}{809} (\bibinfo{year}{2011}), ISSN \bibinfo{issn}{1748-3395}.

\bibitem[{\citenamefont{Platz et~al.}(2012)\citenamefont{Platz, Forchheimer,
  Thol\'{e}n, and Haviland}}]{Platz2012a}
\bibinfo{author}{\bibfnamefont{D.}~\bibnamefont{Platz}},
  \bibinfo{author}{\bibfnamefont{D.}~\bibnamefont{Forchheimer}},
  \bibinfo{author}{\bibfnamefont{E.~A.} \bibnamefont{Thol\'{e}n}},
  \bibnamefont{and} \bibinfo{author}{\bibfnamefont{D.~B.}
  \bibnamefont{Haviland}}, \bibinfo{journal}{Nanotechnology}
  \textbf{\bibinfo{volume}{23}}, \bibinfo{pages}{265705}
  (\bibinfo{year}{2012}), ISSN \bibinfo{issn}{1361-6528}.

\bibitem[{\citenamefont{Platz et~al.}(2013{\natexlab{a}})\citenamefont{Platz,
  Forchheimer, Thol\'{e}n, and Haviland}}]{Platz2013}
\bibinfo{author}{\bibfnamefont{D.}~\bibnamefont{Platz}},
  \bibinfo{author}{\bibfnamefont{D.}~\bibnamefont{Forchheimer}},
  \bibinfo{author}{\bibfnamefont{E.~A.} \bibnamefont{Thol\'{e}n}},
  \bibnamefont{and} \bibinfo{author}{\bibfnamefont{D.~B.}
  \bibnamefont{Haviland}}, \bibinfo{journal}{Nature Communications}
  \textbf{\bibinfo{volume}{4}}, \bibinfo{pages}{1360}
  (\bibinfo{year}{2013}{\natexlab{a}}), ISSN \bibinfo{issn}{2041-1723}.

\bibitem[{\citenamefont{Platz et~al.}(2013{\natexlab{b}})\citenamefont{Platz,
  Forchheimer, Thol\'{e}n, and Haviland}}]{Platz2012b}
\bibinfo{author}{\bibfnamefont{D.}~\bibnamefont{Platz}},
  \bibinfo{author}{\bibfnamefont{D.}~\bibnamefont{Forchheimer}},
  \bibinfo{author}{\bibfnamefont{E.~A.} \bibnamefont{Thol\'{e}n}},
  \bibnamefont{and} \bibinfo{author}{\bibfnamefont{D.~B.}
  \bibnamefont{Haviland}}, \bibinfo{journal}{Beilstein Journal of
  Nanotechnology} \textbf{\bibinfo{volume}{4}}, \bibinfo{pages}{45}
  (\bibinfo{year}{2013}{\natexlab{b}}), ISSN \bibinfo{issn}{2190-4286}.

\bibitem[{\citenamefont{Rodriguez and Garc\'{\i}a}(2002)}]{Rodrguez2002}
\bibinfo{author}{\bibfnamefont{T.~R.} \bibnamefont{Rodriguez}}
  \bibnamefont{and}
  \bibinfo{author}{\bibfnamefont{R.}~\bibnamefont{Garc\'{\i}a}},
  \bibinfo{journal}{Applied Physics Letters} \textbf{\bibinfo{volume}{80}},
  \bibinfo{pages}{1646} (\bibinfo{year}{2002}), ISSN \bibinfo{issn}{00036951}.

\bibitem[{\citenamefont{Melcher et~al.}(2007)\citenamefont{Melcher, Hu, and
  Raman}}]{Melcher2007}
\bibinfo{author}{\bibfnamefont{J.}~\bibnamefont{Melcher}},
  \bibinfo{author}{\bibfnamefont{S.}~\bibnamefont{Hu}}, \bibnamefont{and}
  \bibinfo{author}{\bibfnamefont{A.}~\bibnamefont{Raman}},
  \bibinfo{journal}{Applied Physics Letters} \textbf{\bibinfo{volume}{91}},
  \bibinfo{pages}{053101} (\bibinfo{year}{2007}), ISSN
  \bibinfo{issn}{00036951}.

\bibitem[{\citenamefont{Platz et~al.}(2008)\citenamefont{Platz, Thol\'{e}n,
  Pesen, and Haviland}}]{Platz2008}
\bibinfo{author}{\bibfnamefont{D.}~\bibnamefont{Platz}},
  \bibinfo{author}{\bibfnamefont{E.~A.} \bibnamefont{Thol\'{e}n}},
  \bibinfo{author}{\bibfnamefont{D.}~\bibnamefont{Pesen}}, \bibnamefont{and}
  \bibinfo{author}{\bibfnamefont{D.~B.} \bibnamefont{Haviland}},
  \bibinfo{journal}{Applied Physics Letters} \textbf{\bibinfo{volume}{92}},
  \bibinfo{pages}{153106} (\bibinfo{year}{2008}), ISSN
  \bibinfo{issn}{00036951}.

\bibitem[{\citenamefont{Platz et~al.}(2010)\citenamefont{Platz, Thol\'{e}n,
  Hutter, von Bieren, and Haviland}}]{Platz2010}
\bibinfo{author}{\bibfnamefont{D.}~\bibnamefont{Platz}},
  \bibinfo{author}{\bibfnamefont{E.~A.} \bibnamefont{Thol\'{e}n}},
  \bibinfo{author}{\bibfnamefont{C.}~\bibnamefont{Hutter}},
  \bibinfo{author}{\bibfnamefont{A.~C.} \bibnamefont{von Bieren}},
  \bibnamefont{and} \bibinfo{author}{\bibfnamefont{D.~B.}
  \bibnamefont{Haviland}}, \bibinfo{journal}{Ultramicroscopy}
  \textbf{\bibinfo{volume}{110}}, \bibinfo{pages}{573} (\bibinfo{year}{2010}),
  ISSN \bibinfo{issn}{1879-2723}.

\bibitem[{\citenamefont{Derjaguin et~al.}(1975)\citenamefont{Derjaguin, Muller,
  and Toporov}}]{DERJAGUIN1975}
\bibinfo{author}{\bibfnamefont{B.~V.} \bibnamefont{Derjaguin}},
  \bibinfo{author}{\bibfnamefont{V.~M.} \bibnamefont{Muller}},
  \bibnamefont{and} \bibinfo{author}{\bibfnamefont{Y.~P.}
  \bibnamefont{Toporov}}, \bibinfo{journal}{Journal of Colloid and Interface
  Science} \textbf{\bibinfo{volume}{53}}, \bibinfo{pages}{314}
  (\bibinfo{year}{1975}), ISSN \bibinfo{issn}{00219797}.

\bibitem[{\citenamefont{Luan and Robbins}(2005)}]{Luan2005}
\bibinfo{author}{\bibfnamefont{B.}~\bibnamefont{Luan}} \bibnamefont{and}
  \bibinfo{author}{\bibfnamefont{M.~O.} \bibnamefont{Robbins}},
  \bibinfo{journal}{Nature} \textbf{\bibinfo{volume}{435}},
  \bibinfo{pages}{929} (\bibinfo{year}{2005}), ISSN \bibinfo{issn}{1476-4687}.

\bibitem[{\citenamefont{Brandrup et~al.}(2005)\citenamefont{Brandrup,
  {Immergut, Edmund H. Grulke}, Abe, and Bloch}}]{Brandrup2005}
\bibinfo{author}{\bibfnamefont{J.}~\bibnamefont{Brandrup}},
  \bibinfo{author}{\bibfnamefont{E.~A.} \bibnamefont{{Immergut, Edmund H.
  Grulke}}}, \bibinfo{author}{\bibfnamefont{A.}~\bibnamefont{Abe}},
  \bibnamefont{and} \bibinfo{author}{\bibfnamefont{D.~R.} \bibnamefont{Bloch}},
  \emph{\bibinfo{title}{{Polymer Handbook}}} (\bibinfo{publisher}{Wiley \&
  Sons}, \bibinfo{year}{2005}), \bibinfo{edition}{4th} ed., ISBN
  \bibinfo{isbn}{9781591248835}.

\bibitem[{\citenamefont{Sahin and Erina}(2008)}]{Sahin2008a}
\bibinfo{author}{\bibfnamefont{O.}~\bibnamefont{Sahin}} \bibnamefont{and}
  \bibinfo{author}{\bibfnamefont{N.}~\bibnamefont{Erina}},
  \bibinfo{journal}{Nanotechnology} \textbf{\bibinfo{volume}{19}},
  \bibinfo{pages}{445717} (\bibinfo{year}{2008}), ISSN
  \bibinfo{issn}{0957-4484}.

\bibitem[{\citenamefont{Platz et~al.}(2013{\natexlab{c}})\citenamefont{Platz,
  Forchheimer, Thol\'{e}n, and Haviland}}]{Platz2012d}
\bibinfo{author}{\bibfnamefont{D.}~\bibnamefont{Platz}},
  \bibinfo{author}{\bibfnamefont{D.}~\bibnamefont{Forchheimer}},
  \bibinfo{author}{\bibfnamefont{E.~A.} \bibnamefont{Thol\'{e}n}},
  \bibnamefont{and} \bibinfo{author}{\bibfnamefont{D.~B.}
  \bibnamefont{Haviland}} (\bibinfo{year}{2013}{\natexlab{c}}),
  \eprint{1301.7340}.

\bibitem[{\citenamefont{Higgins et~al.}(2006)\citenamefont{Higgins, Proksch,
  Sader, Polcik, {Mc Endoo}, Cleveland, and Jarvis}}]{Higgins2006}
\bibinfo{author}{\bibfnamefont{M.~J.} \bibnamefont{Higgins}},
  \bibinfo{author}{\bibfnamefont{R.}~\bibnamefont{Proksch}},
  \bibinfo{author}{\bibfnamefont{J.~E.} \bibnamefont{Sader}},
  \bibinfo{author}{\bibfnamefont{M.}~\bibnamefont{Polcik}},
  \bibinfo{author}{\bibfnamefont{S.}~\bibnamefont{{Mc Endoo}}},
  \bibinfo{author}{\bibfnamefont{J.~P.} \bibnamefont{Cleveland}},
  \bibnamefont{and} \bibinfo{author}{\bibfnamefont{S.~P.}
  \bibnamefont{Jarvis}}, \bibinfo{journal}{Review of Scientific Instruments}
  \textbf{\bibinfo{volume}{77}}, \bibinfo{pages}{013701}
  (\bibinfo{year}{2006}), ISSN \bibinfo{issn}{00346748}.

\bibitem[{\citenamefont{Thol\'{e}n et~al.}(2011)\citenamefont{Thol\'{e}n,
  Platz, Forchheimer, Schuler, Thol\'{e}n, Hutter, and Haviland}}]{Tholen2011}
\bibinfo{author}{\bibfnamefont{E.~A.} \bibnamefont{Thol\'{e}n}},
  \bibinfo{author}{\bibfnamefont{D.}~\bibnamefont{Platz}},
  \bibinfo{author}{\bibfnamefont{D.}~\bibnamefont{Forchheimer}},
  \bibinfo{author}{\bibfnamefont{V.}~\bibnamefont{Schuler}},
  \bibinfo{author}{\bibfnamefont{M.~O.} \bibnamefont{Thol\'{e}n}},
  \bibinfo{author}{\bibfnamefont{C.}~\bibnamefont{Hutter}}, \bibnamefont{and}
  \bibinfo{author}{\bibfnamefont{D.~B.} \bibnamefont{Haviland}},
  \bibinfo{journal}{Review of Scientific Instruments}
  \textbf{\bibinfo{volume}{82}}, \bibinfo{pages}{026109}
  (\bibinfo{year}{2011}), ISSN \bibinfo{issn}{1089-7623}.

\end{thebibliography}
\end{document}